\documentclass[onecolumn,authoryear]{els-mrw} 

\usepackage{amsmath,amssymb,amsfonts,amsthm,makeidx,graphicx}
\usepackage{txfonts}
\usepackage{helvet}
\usepackage[dvipsnames]{xcolor} 
\usepackage{mathtools} 
\usepackage[colorlinks]{hyperref} 
\usepackage{textgreek} 
\usepackage{upgreek} 

\newcommand{\figRef}[1]{\mbox{Figure \ref{fig: #1}}}
\newcommand{\eqRef}[1]{\mbox{Equation \ref{eq: #1}}}

\long\def\symbolfootnote[#1]#2{\begingroup%
\def\thefootnote{\fnsymbol{footnote}}\footnote[#1]{#2}\endgroup} 
\def\aj{AJ}
\def\araa{ARA\&A}
\def\apj{ApJ}
\def\apjl{ApJ}
\def\aap{A\&A}
\def\icarus{Icarus}
\def\jgr{J. Geo. Res.}
\def\mnras{MNRAS}
\def\planss{P\&SS}

\newcommand{\presecContentsEntry}[1]{\noindent\nameref{#1} \hfill \pageref{#1} \newline}
\newcommand{\secContentsEntry}[1]{\noindent\ref{#1}. \nameref{#1} \hfill \pageref{#1} \newline}
\newcommand{\subsecContentsEntry}[1]{\forceindent\ref{#1}. \nameref{#1} \hfill \pageref{#1}\newline}
\newcommand{\forceindent}{\leavevmode{\parindent=2em\indent}}

\newcommand{\glossaryTerm}[1]{`{#1}'}

\newcommand{\boxTextColour}[1]{{\color{Maroon}{#1}}}

\newcommand{\horizontalLine}[1]{\noindent\textcolor{darkgray}{\rule{\textwidth}{1pt} }}

\newenvironment{absolutelynopagebreak}
  {\par\nobreak\vfil\penalty0\vfilneg
   \vtop\bgroup}
  {\par\xdef\tpd{\the\prevdepth}\egroup
   \prevdepth=\tpd}

\mathchardef\shorthyphen="2D                                            
\newcommand{\lSun}{\; {\rm L_\odot}}                            
\newcommand{\au}{\; {\rm au}}                                           
\newcommand{\pc}{\; {\rm pc}}                                           
\newcommand{\mm}{\; {\rm mm}}                                                     
\newcommand{\um}{\; {\rm \upmu m}}                                        
\newcommand{\myr}{\; {\rm Myr}}                                         
\newcommand{\K}{\; {\rm K}}                                                     
\newcommand{\HD}{{\rm HD} \;}                                           

\newcommand{\rHill}{r_{\rm H}}                                        
\newcommand{\tBB}{T_{\rm BB}}                                         
\newcommand{\rBB}{r_{\rm BB}}                                         
\newcommand{\rDust}{r_{\rm d}}                                        
\newcommand{\lStar}{L_*}                                              
\newcommand{\mStar}{M_*}                                              
\newcommand{\mPlt}{m_{\rm p}}                                         
\newcommand{\mDust}{m_{\rm d}}                                        
\newcommand{\aPlt}{a_{\rm p}}                                         
\newcommand{\ePlt}{e_{\rm p}}                                         
\newcommand{\eMax}{e_{\rm max}}                                       
\newcommand{\fGrav}{\boldsymbol{F}_{\rm grav}}                        
\newcommand{\rHat}{\boldsymbol{\hat{r}}}                              
\newcommand{\aRes}{a_{\rm res}}                                       
\newcommand{\QDStar}{Q_{\rm D}^*}                                     

\begin{document}

\chapter{Debris disks around main-sequence stars}\label{chap1}

\author[1]{Tim D. Pearce}%

\address[1]{\orgdiv{Department of Physics}, \orgname{University of Warwick}, \orgaddress{UK}}%


\maketitle

\begin{abstract}[Abstract]

\glossaryTerm{Debris disks} are collections of small bodies around stars, such as the Asteroid Belt and Kuiper Belt in our Solar System. These disks are composed of objects smaller than planets, including asteroids, comets, dust, and dwarf planets. We detect debris disks around a significant fraction of stars, and these disks appear to be common components of planetary systems. Extrasolar debris disks have a broad range of locations, shapes and features. This chapter provides an introduction to debris disks around main-sequence stars. It summarises our understanding of the field, and covers a wide range of concepts from observations and theory. It describes how we detect extrasolar debris disks, what we see, and what these observations tell us. It also describes how debris disks evolve, and how they interact with planets. The chapter concludes by discussing several unsolved questions in debris-disk science.

\end{abstract}


\vspace{5mm}

\begin{keywords}
    Debris disks, circumstellar disks, circumstellar dust, circumstellar gas, exozodiacal dust, planetesimals, planetary system formation, planetary system evolution, collisional processes
\end{keywords}

\vspace{5mm}

\horizontalLine

\noindent \textbf{Contents}

\presecContentsEntry{sec: learningObjectives}
\presecContentsEntry{presec: glossary}
\presecContentsEntry{presec: nomenclature}
\secContentsEntry{sec: introduction}
\secContentsEntry{sec: obs}
\subsecContentsEntry{subsec: obsDust}
\subsecContentsEntry{subsec: obsGas}
\secContentsEntry{sec: theory}
\subsecContentsEntry{subsec: theoryOriginOfDebrisDisks}
\subsecContentsEntry{subsec: theoryWhyDifferentAtDifferentWavelengths}
\subsecContentsEntry{subsec: theorySizeDistAndCollCascade}
\subsecContentsEntry{subsec: theoryPlanetDebrisInteractions}
\subsecContentsEntry{subsec: theoryHotAndWarmDust}
\secContentsEntry{sec: unsolvedProblems}
\secContentsEntry{sec: conclusions}
\horizontalLine

\vspace{5mm}

\section*{Learning objectives}
\label{sec: learningObjectives}

\vspace{1mm}

\noindent By the end of this chapter, you should understand:

\vspace{1mm}

\begin{itemize}
    \item What a debris disk is
    \item How we observe debris around main-sequence stars
    \item Where debris is located around stars
    \item How debris evolves through dynamical forces and collisions
    \item How planets interact with debris
    \item What the major questions are in debris-disk science
\end{itemize}

\vspace{5mm}

\begin{glossary}[Glossary]
\label{presec: glossary}

    \term{Apoastron (or apastron)} The point on an eccentric orbit around a star which is furthest from the star.
    
    \term{Back scattering} Where light scattered by a dust grain is deflected by almost ${180^\circ}$ from its incident angle.

    \term{Blackbody} A theoretical body that is a perfect absorber and emitter of radiation.

    \term{Blackbody radius} The distance a blackbody would have to lie from a star to have a certain temperature.

    \term{Blowout size} The size of a dust grain just small enough to be blown away from a star by radiation pressure or stellar winds.
    
    \term{Catastrophic collision} A collision where the largest surviving fragment has less than half the mass of the original body.

    \term{Clump} An overdensity that does not extend all the way around the star.

    \term{Cold dust} Dust in the outer regions of planetary systems, with typical temperatures of order ${30\K}$.

    \term{Collisional cascade} Bodies colliding and breaking up into smaller bodies, which then collide and break into even smaller bodies, etc.
    
    \term{Continuum} Component of an SED that is relatively smooth, as opposed to the sharp features at spectral lines.

    \term{Coronagraph} A device that blocks out starlight, allowing fainter objects to be resolved.
    
    \term{Debris} Solid objects smaller than planets, such as asteroids, comets, dust and dwarf planets.

    \term{Debris disk} A population of debris bodies, such as the Asteroid Belt and Kuiper Belt in the Solar System.

    \term{Exozodi} Warm dust near the habitable zone in extrasolar systems, which may be analogous to Zodiacal dust in the Solar System.

    \term{Extreme debris disk} A debris disk with an unusually high luminosity, typically above ${1\%}$ of the star's luminosity.

    \term{Forward scattering} Where light scattered by a dust grain is deflected only slightly from its incident angle.

    \term{Gamma factor} The ratio of dust's actual location to its blackbody radius.

    \term{Halo} A population of small grains on eccentric orbits, which extend out beyond the main debris disk.

    \term{Hill radius} For a body orbiting a star, the distance around the body where its gravity dominates over that of the star.

    \term{Hot dust} Dust very close to stars, with temperatures of order ${1000\K}$. Also known as a hot exozodi.

    \term{Hot exozodi} Another name for hot dust.

    \term{Hybrid disk} A young disk with large amounts of gas but relatively small amounts of dust. It may contain primordial gas but collisional dust, and be transitioning from a protoplanetary disk into a debris disk.

    \term{Infrared excess} Flux from dust, which makes a system brighter at infrared wavelengths than would be expected from the star alone.

    \term{Interferometer} Several telescopes combined together to produce higher-resolution images.
    
    \term{Main-sequence star} A star that fuses hydrogen into helium in its core. Most stars, including the Sun, are in this phase of their lives.    

    \term{Mean-motion resonance (MMR)} A dynamical interaction that occurs if the orbital periods of bodies are near simple fractions of each other.

    \term{Metallicity} The abundance of elements other than hydrogen and helium in an object.

    \term{Nominal MMR location} The semimajor axis where two bodies in the ${(p+q):p}$ MMR have an orbital-period ratio of exactly ${(p+q):p}$.

    \term{Optically thin} Describes a medium or object where almost all incident light is transmitted through it.

    \term{Order (of MMR)} The value of $q$ in the ${(p+q):p}$ MMR.
    
    \term{Periastron} The point on an eccentric orbit around a star which is closest to the star.

    \term{Phase function} The relationship between the scattered-light brightness and the scattering angle.

    \term{Planetary system} At least one star, plus any planets and other material that formed from its protoplanetary disk.
        
    \term{Planetesimal} A solid object which is smaller than a planet but larger than a dust grain, such as an asteroid.

    \term{Polarisation} The direction in which an electromagnetic wave's electric field oscillates. 

    \term{Poynting-Robertson (P-R) drag} A drag force arising from photons hitting a moving dust grain.

    \term{Protoplanetary disk} A massive disk of gas and dust surrounding a young star, from which planets and debris may form.
    
    \term{Radiation pressure} A force arising from photons hitting a dust grain, which acts in the direction away from the star.

    \term{Scattered light} Light reflected off dust grains.

    \term{Scattering (dynamical)} A short-term dynamical interaction, occurring when two bodies make a close approach to each other.

    \term{Scattering angle} The angle between the incident photon's path and the scattered photon's path.

    \term{Secular interaction} A long-term dynamical interaction, occurring over timescales much longer than orbital periods.

    \term{Spectral-energy distribution (SED)} A plot of the relationship between flux and wavelength.
    
    \term{Spectral lines} Narrow spikes or dips in an SED, which correspond to specific quantum processes in atoms or molecules.

    \term{Stirring} A dynamical process that excites debris orbits, causing debris bodies to collide destructively.

    \term{Thermal emission} Photons radiated as heat from objects.
    
    \term{Transition disk} A disk of gas and dust around a young star, with a cavity at its inner region. May be transitioning from a protoplanetary disk to a debris disk. 

    \term{Trojan} Objects on a similar orbit to a planet, in a 1:1 MMR and located about $60^\circ$ ahead of or behind the planet.

    \term{Volatiles} Substances that easily vaporize.

    \term{von Zeipel-Kozai-Lidov mechanism} A secular interaction, sometimes called the Kozai mechanism, where a body oscillates between a low-eccentricity orbit with high inclination, and a high-eccentricity orbit with low inclination.
    
    \term{Warm dust} Dust located near the habitable regions of planetary systems, with temperatures of order ${300\K}$.

    \term{Zodiacal dust} Interplanetary dust in the Solar System.

\end{glossary}

\begin{glossary}[Nomenclature]
\label{presec: nomenclature}

\begin{tabular}{@{}lp{34pc}@{}}
\term{ALMA} & The Atacama Large Millimeter/submillimeter Array\\
\term{GPI} & The Gemini Planet Imager, on the Gemini South Telescope\\
\term{\textit{HST}} & The \textit{Hubble Space Telescope}\\
\term{\textit{JWST}} & The \textit{James Webb Space Telescope}\\
\term{MMR} & Mean-motion resonance\\
\term{P-R drag} & Poynting-Robertson drag\\
\term{SED} & Spectral-energy distribution\\
\term{SPHERE} & The Spectro-Polarimetric High-contrast Exoplanet REsearch instrument, on the VLT\\
\term{VLT} & The Very Large Telescope\\
\end{tabular}
\end{glossary}

\vspace{5mm}

\horizontalLine

\section{Introduction: what are debris disks?}
\label{sec: introduction}

Our Solar System is a \glossaryTerm{planetary system}, which means a system of bodies orbiting one or more stars. It contains the Sun, planets and moons, but also many small objects, such as asteroids, comets, dwarf planets and dust. Collectively, these small objects are called \glossaryTerm{debris}. Debris exists almost everywhere in the Solar System, from Sun-grazing comets in the inner regions, to interplanetary dust, the Asteroid Belt, Kuiper Belt, and long-period comets further out. Debris is also not unique to the Solar System; we detect debris around other stars too. We think that all stars probably host debris in some form, some of which is similar to debris in the Solar System, and some different. \figRef{debrisAcrossSystemsSchematic} shows some types of debris detected so far. 

\begin{figure}[h]
\centering
\includegraphics[width=12cm]{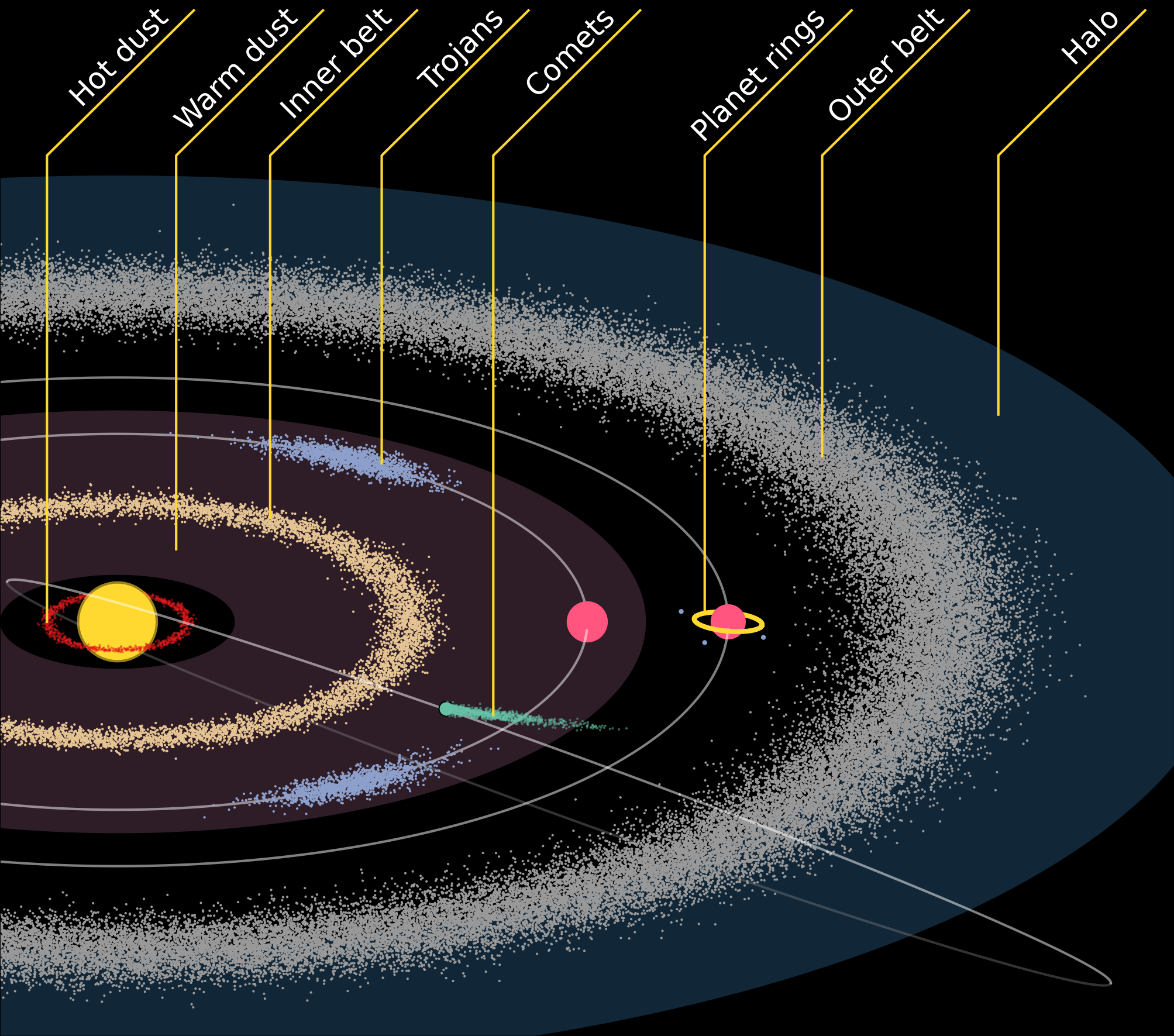}
\caption{Some of the debris found in planetary systems. The star is the large yellow circle, and the two smaller circles are planets.}
\label{fig: debrisAcrossSystemsSchematic}
\end{figure}

A collection of debris in a planetary system is called a \glossaryTerm{debris disk}. This term can either mean \textit{all} the debris in a planetary system, or it can mean \textit{individual} debris populations, like the Asteroid Belt or Kuiper Belt. The box below provides a very brief outline of a debris disk and how it evolves, to bear in mind as you read this chapter. Each of the concepts mentioned will be explained later.

\begin{absolutelynopagebreak}
\begin{BoxTypeA}[chap1:box1]{Key concept: \boxTextColour{outline of a debris disk}}

\noindent To picture a debris disk, imagine a collection of bodies in a planetary system. These range in size from small dust grains up to asteroids, and possibly dwarf planets. These bodies break apart over time due to violent collisions and other processes. This fragmentation releases dust and gas, which we can detect in extrasolar systems. As the debris bodies get smaller they experience different forces, and when they get small enough they are removed from the system by radiation forces. This means that debris disks lose mass over time.


\end{BoxTypeA}
\end{absolutelynopagebreak}

This chapter focuses on debris around \glossaryTerm{main-sequence} stars, which are stars in the mid-stage of their lives. Other chapters cover debris at other stages in a star's life. This chapter also focuses on extrasolar debris, with other chapters covering Solar System bodies. The first part of this chapter describes observations of extrasolar debris disks (\mbox{Section \ref{sec: obs}}), and the second part describes our theoretical understanding of debris (\mbox{Section \ref{sec: theory}}). The final part discusses some unanswered questions in debris science (\mbox{Section \ref{sec: unsolvedProblems}}). Throughout this chapter, new terms are highlighted in quotation marks, and these terms are listed in the glossary.

\horizontalLine

\section{Observations of extrasolar debris}
\label{sec: obs}

Many people know that we can detect extrasolar planets. We also detect extrasolar debris, with the first such detection made in the 1980s around the star Vega \citep{Aumann1984}. Since then, we have detected debris around at least 600 main-sequence stars within ${200\pc}$ of Earth (\mbox{Matthews et al., in prep.}). Surveys find debris around ${20\%}$ of main-sequence stars, and this ${20\%}$ almost certainly underestimates the true fraction of stars with debris disks, because modern instruments are only sensitive to the brightest disks \citep{Wyatt2008}. For example, the Solar System's Asteroid and Kuiper Belts, were they located around another star, would be too faint to detect with current technology.

There are several techniques for detecting extrasolar debris. These range from individual flux measurements, which hint that debris is present, to resolved images that show debris structures in exquisite detail. \figRef{galleryOfDisks} shows several extrasolar debris disks, resolved with different instruments and showing different types of debris. Each of these disks have distinct features, which will be discussed throughout this chapter. This section explains how we detect debris, which is done by observing dust and gas, and what these observations show.

\begin{figure}[h]
\centering
\includegraphics[width=16.5cm]{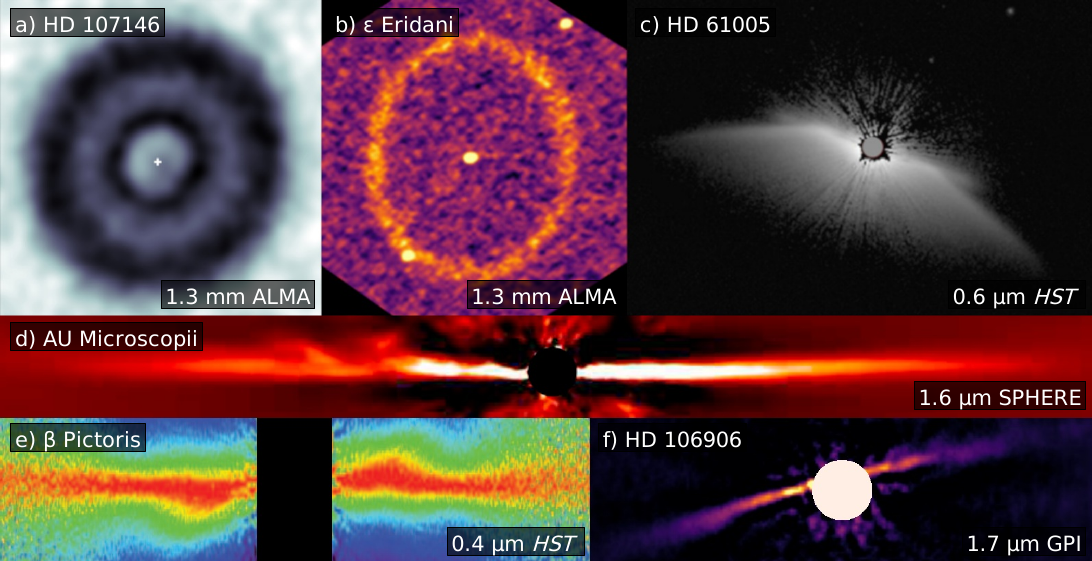}
\caption{Observations of dust around main-sequence stars. These are \glossaryTerm{cold} debris disks in the outer regions of planetary systems, at 10s or 100s of au from their stars. Each panel shows a different system, with the star at the centre. Each panel gives the system name in the top left, and the image wavelength followed by the instrument name in the bottom right. In images \textit{c}) to \textit{f}), coronagraphs were used to block out the star, so no data are available inside the filled areas. References: \mbox{\textit{a}) \citet{Marino2018}}, \mbox{\textit{b}) \citet{Booth2023}}, \mbox{\textit{c}) \citet{Schneider2014}}, \mbox{\textit{d}) \citet{Boccaletti2018}}, \mbox{\textit{e}) \citet{Golimowski2006}}, \mbox{\textit{f}) \citet{Hom2024}}.}
\label{fig: galleryOfDisks}
\end{figure}

\subsection{Dust observations}
\label{subsec: obsDust}

Most of our information about extrasolar debris disks comes from observations of dust. This is because dust has a large surface area, so is brighter and easier to detect than larger objects. When dust gets illuminated by a star, it emits photons in two main ways: \glossaryTerm{scattered light}, and \glossaryTerm{thermal emission}. These processes let us detect extrasolar dust.

\begin{BoxTypeA}[chap1:box1]{Key concept: \boxTextColour{dust is seen in scattered light and thermal emission}}

\vspace{1mm}
\begin{itemize}
    \item \textbf{Scattered light.} This is where light from a star gets reflected off a dust grain.

    \vspace{1mm}
    
    \item \textbf{Thermal emission.} This is where a grain radiates heat as photons.
    
\end{itemize}
\vspace{1mm}

\end{BoxTypeA}


Scattered light and thermal emission emit photons at different wavelengths. For scattered light, emitted photons have similar wavelengths to the incident photons.
This means that dust seen in scattered light has a similar spectrum to the star. For thermal emission, the emitted wavelengths depend on dust temperature, with colder grains emitting at longer wavelengths. The temperature depends on distance to the star, with more-distant dust being colder. This temperature is often used to categorise dust:

\vspace{1mm}
\begin{itemize}
    \item \textbf{\glossaryTerm{Cold dust}:} dust at temperatures of order ${30 \K}$. Located in the outer regions of planetary systems, at 10s to 100s \mbox{of au} from stars.

    \vspace{1mm}
    
    \item \textbf{\glossaryTerm{Warm dust}:} temperatures of order ${300 \K}$. Distances of order ${1\au}$, which is near the habitable zone.

    \vspace{1mm}

    \item \textbf{\glossaryTerm{Hot dust}:} temperatures of order ${1000 \K}$. Located very close to stars, at ${0.1\au}$ or less.
    
\end{itemize}
\vspace{1mm}

Temperature is not the only factor that sets the wavelengths of dust emission. Another important factor is grain size, with smaller grains emitting at shorter wavelengths.

\begin{BoxTypeA}[chap1:box1]{Key concept: \boxTextColour{observations detect grains with sizes similar to the observing wavelength}}

\noindent When we observe dust, we detect grains with sizes similar to the observing wavelength. This is due to a combination of two effects:

\vspace{1mm}
\begin{itemize}
    \item \textbf{Grains are inefficient at emitting photons with wavelengths larger than the grain size.}

    \vspace{1mm}

    \item \textbf{Smaller debris is more numerous.} The smallest dust dominates a debris disk's surface area, so these grains can absorb and emit more photons.
    
\end{itemize}
\vspace{1mm}

\noindent These two effects mean that emission at one wavelength is mainly from grains with sizes around that wavelength. The instruments that detect extrasolar debris mainly operate between sub-micron and millimetre wavelengths, because debris is too faint at wavelengths longer than this. This means that we cannot directly detect extrasolar debris larger than a few centimetres.

\end{BoxTypeA}

\subsubsection{Spectral-energy distributions (SEDs)}
\label{subsec: obsSEDs}

The first detections of extrasolar debris were made by analysing \glossaryTerm{spectral-energy distributions} (SEDs), which are relationships between a system's flux and wavelength. These SEDs made it clear that debris exists in planetary systems, even though we could not see it directly at the time. \figRef{betaPicSED} shows the SED of the \mbox{\textbeta\;Pictoris} system, which hosts at least two dust populations. 

\begin{figure}[h]
\centering
\includegraphics[width=14cm]{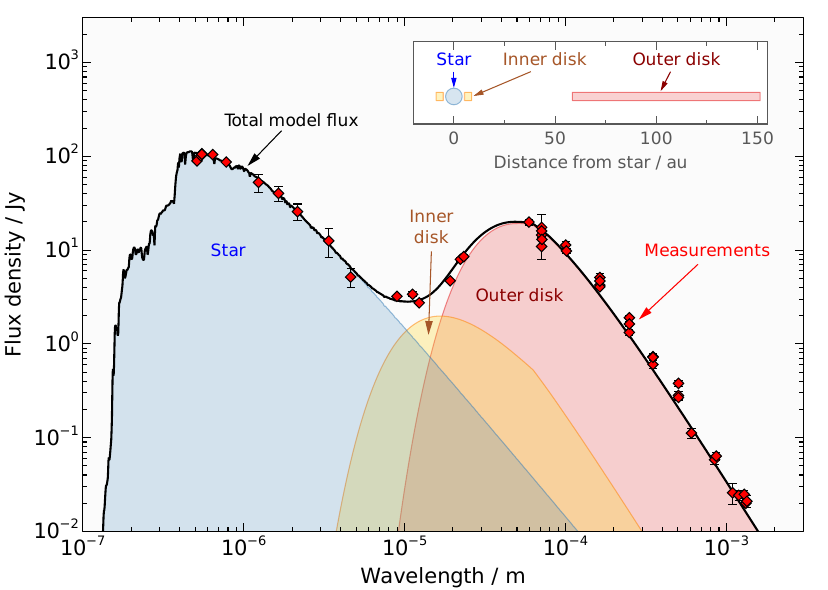}
\caption{SED of the \mbox{\textit{\textbeta} Pictoris} system. This system has an \glossaryTerm{infrared excess}, indicating the presence of debris. Red diamonds show measured fluxes, and lines are models. The star flux (blue line) dominates at shorter wavelengths, but cannot reproduce the data at longer, infrared wavelengths. This infrared excess demonstrates that two debris populations are present: a colder outer disk, and a warmer inner disk. The fluxes from these populations are the red and orange lines respectively. Most of the debris flux comes from the outer disk, but that disk does not produce enough flux around \mbox{10 {\textmu}m} to reproduce observations, which is why the inner-disk model is also needed. The black line is the combined flux from the star and debris models, which reproduces the observational data. The inset shows the locations of debris in the system. The data and modelling are described in \cite{Yelverton2019}.}
\label{fig: betaPicSED}
\end{figure}

Individual objects have different SEDs, depending on their temperature and composition. A star's SED is similar to that of a \glossaryTerm{blackbody}, which is a theoretical body that is a perfect absorber and emitter of radiation. A star's SED peaks at wavelengths that depends on its temperature, and declines at longer wavelengths. The SEDs of isolated stars are known very accurately, so if a system's SED deviates from the expected star SED, then other objects must be present. This can be used to detect dust in planetary systems. The dust causes the measured flux (red diamonds on \figRef{betaPicSED}) to significantly deviate from the stellar SED (blue line). Since dust is colder than stars, the thermal emission from dust peaks at longer wavelengths. This peak is often at infrared wavelengths, so systems with dust are sometimes said to have \glossaryTerm{infrared excesses}. Scattered light from dust also contributes, but this component is difficult to detect in SEDs because it is very similar to the star, and much fainter.

The dust quantity and location can be calculated from SEDs. Dust quantity is related to its brightness, which can be calculated as the total system flux minus the stellar flux. Dust in debris disks is \glossaryTerm{optically thin}, meaning it has a low optical depth and hence no dust is hidden from view. This means the dust brightness can be converted directly into dust quantity, by making some assumptions about its optical properties (e.g. \mbox{Equation 5} in \citealt{Krivov2021}). 

Dust location can also be inferred from SEDs. The SED of dust is similar to a blackbody, and peaks at a wavelength that depends on its temperature. Fitting a blackbody to the dust SED yields a temperature $\tBB$, which can be converted to a \glossaryTerm{blackbody radius} $\rBB$ using

\begin{equation}
\rBB = 1\au \; \left(\frac{\tBB}{278\K}\right)^{-2} \left(\frac{\lStar}{\lSun}\right)^{1/2},
\label{eq: rBBFromTBB}
\end{equation}

\noindent where $\lStar$ and $\lSun$ are the luminosity of the star and Sun respectively. However, the blackbody radius $\rBB$ often does not correspond to the actual location of the dust, $\rDust$, because grains are not perfect blackbodies. To account for this, we multiply $\rBB$ by the empirical \glossaryTerm{Gamma Factor} $\Gamma$:

\begin{equation}
\rDust = \Gamma \rBB.
\label{eq: rDustFromRBB}
\end{equation}

\noindent The value of $\Gamma$ varies between about 1 and 4, and depends on star luminosity \citep{Pawellek2015}. 

SEDs can identify cold dust in the outer regions of systems, and also warmer dust closer to stars. The example on \figRef{betaPicSED} shows a system with both cold and warm dust, where two blackbodies are required to fit the observations. These two blackbodies peak at different wavelengths, implying that the system hosts dust at two different temperatures (and hence two different locations).

\subsubsection{Resolved images of cold dust}
\label{subsec: obsResolvedImagesOfColdDust}

SEDs are useful for constraining the dust quantity, and its rough location. However, SEDs are often less informative than resolved images, because resolved images reveal the shape and extent of the dust distribution. \figRef{galleryOfDisks} shows some resolved images, and about 200 debris disks have now been resolved\footnote{Catalogues of resolved debris disks are available at \href{https://www.circumstellardisks.org/}{circumstellardisks.org/} and \href{https://www.astro.uni-jena.de/index.php/theory/catalog-of-resolved-debris-disks.html}{astro.uni-jena.de/index.php/theory/catalog-of-resolved-debris-disks.html}.}. Some resolved images are made using one telescope, whilst others are \glossaryTerm{interferometric}, where multiple telescopes are combined to produce higher-resolution images.

Resolved observations show dust in either scattered light or thermal emission, depending on the observing wavelength. Scattered-light observations have relatively short wavelengths, typically around ${1\um}$, which is where stars are brightest (\figRef{betaPicSED}). This is because scattered photons have similar wavelengths to incident photons, so scattered light from dust peaks at the same wavelengths as the star. Since the star is orders of magnitude more luminous than dust, scattered-light observations often require \glossaryTerm{coronagraphs}, which block out starlight to reveal fainter debris. Panels \textit{c} to \textit{f} of \figRef{galleryOfDisks} show scattered-light observations, with coronagraphs marked by the filled areas. Since coronagraphs obscure the region around the star, scattered-light observations tend to resolve cold dust in the outer regions of planetary systems.

Thermal-emission observations usually have longer wavelengths, typically between ${10\um}$ and ${1\mm}$. For cold and warm dust, this is where the dust's SED peaks (\figRef{betaPicSED}). At these wavelengths dust is often much brighter than the star, so thermal observations generally do not require coronagraphs. Panels \textit{a} and \textit{b} of \figRef{galleryOfDisks} show thermal observations; the star is not visible on panel \textit{a}, and does not outshine the dust on panel \textit{b}. Like scattered light, thermal images between ${10\um}$ and ${1\mm}$ tend to show cold dust in the outer regions of systems. They can also detect warm dust closer to stars, but cold dust is often more plentiful, so is usually brighter.

The main difference between thermal and scattered-light observations is the type of dust they probe. Thermal-emission observations have longer wavelengths, so probe larger grains. Scattered-light observations have shorter wavelengths, so probe smaller grains. Comparing scattered-light and thermal-emission images reveals that debris disks look different at different wavelengths. Typically, disks appear more compact at longer wavelengths (thermal emission), and more extended at shorter wavelengths (scattered light). This means that dust grains of different sizes have different spatial distributions. \figRef{fomalhautAtDifferentWavelengths} shows an example, where one disk is resolved at three different wavelengths, revealing three different shapes. 

\begin{figure}[h]
\centering
\includegraphics[width=16.5cm]{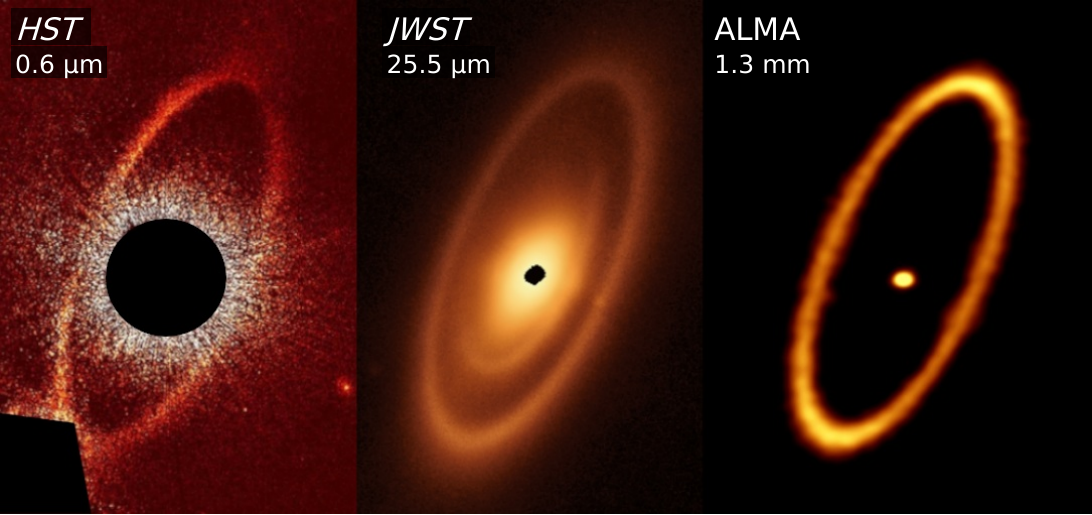}
\caption{Dust around the star Fomalhaut, imaged at three different wavelengths. The main dust ring (visible on all three panels) is slightly elliptical, so the star is offset from its centre. Left panel: \textit{HST} image centred on a wavelength \mbox{0.6 {\textmu}m}, with a coronagraph to mask the star \citep{Kalas2013}. Middle panel: \textit{JWST} image at \mbox{25.5 {\textmu}m}, revealing the main ring and also inner dust, which is not visible at other wavelengths \citep{Gaspar2023}. Right panel: ALMA image at \mbox{1.3 mm}, where the main ring and star are detected \citep{MacGregor2017}.}
\label{fig: fomalhautAtDifferentWavelengths}
\end{figure}

The reason that disks look different at different wavelengths is that grains of different sizes are affected by different forces. This will be explained in \mbox{Section \ref{subsec: theoryWhyDifferentAtDifferentWavelengths}}, but generally, large grains are mainly affected by gravity, whilst small grains are also affected by other forces. Thermal observations of large grains should therefore trace the distribution of \glossaryTerm{planetesimals}, which are objects smaller than planets but larger than dust, and which are thought to be the source of dust in debris disks. We cannot detect planetesimals directly, but they should be co-located with large grains, because they both experience similar forces. Conversely, small grains can have very different orbits. They can form \glossaryTerm{halos}, which are structures extending beyond the planetesimal belt, comprising small grains which are driven onto wide orbits by non-gravitational forces. \figRef{galleryOfDisks}\textit{c} shows a halo in scattered light, where the halo is the wide, triangular structure below and to the left of the star.

\begin{BoxTypeA}[chap1:box1]{Key concept: \boxTextColour{debris disks look different at different wavelengths}}

\noindent Grains of different sizes have different spatial distributions, so debris disks look different at different wavelengths. Disks are usually more compact at longer wavelengths and more extended at shorter wavelengths.

\end{BoxTypeA}

There are also other differences between scattered-light and thermal-emission observations. Scattered-light images can be used to measure \glossaryTerm{polarisation}, which is the orientation of the light's electric-field oscillation. Light from stars is unpolarised, meaning it oscillates in all directions, but scattered light is polarised, meaning it preferentially oscillates in certain directions. Polarisation measurements let us constrain various dust properties, including the composition, size and porosity of dust grains. Scattered-light observations also depend on the \glossaryTerm{phase function}, which is the relationship between brightness and \glossaryTerm{scattering angle} (the angle between the incident photon's path and the scattered photon's path). A dust grain does not scatter light equally in all directions; most light is \glossaryTerm{forward scattered}, meaning it is only slightly deflected from its path, and less light is fully reflected by ${180^\circ}$ (\glossaryTerm{back scattered}). This means that disks seen in scattered light tend to look brighter on the side between us and the star, and fainter on the far side. \figRef{galleryOfDisks}\textit{f} shows an example, where the visible part of the disk (top) is closest to us, and the far side (bottom) appears much fainter.

\subsubsection{Warm and hot dust from interferometry}
\label{subsec: obsHotAndWarmDust}

One way to detect warm dust is from its infrared excess, like on \figRef{betaPicSED}. Another method is to use interferometry, which can separate the dust flux from the star flux. This lets us detect dust located very close to stars. Interferometry is used to detect both warm and hot dust, with fluxes around ${1\%}$ of the star's flux.

Warm dust is detected at mid-infrared wavelengths, around ${10\um}$. This dust appears to lie near the habitable zone, and some may be analogous to the \glossaryTerm{Zodiacal dust} between planets in the Solar System. Hence extrasolar warm-dust populations are sometimes called \glossaryTerm{exozodis}, and we detect them around 20 to ${30\%}$ of main-sequence stars \citep{Ertel2020}. The other population is hot dust, which is detected at near-infrared wavelengths around ${1\um}$. This dust appears to have temperatures of at least ${1000\K}$, and is located very close to stars. Populations of hot dust are sometimes called \glossaryTerm{hot exozodis}, and we also detect them around 20 to ${30\%}$ of main-sequence stars \citep{Ertel2020}.

There are no strong correlations between warm and hot dust, with some stars having both populations, some having one, and some having neither. Warm dust \textit{is} strongly correlated with cold dust, but hot dust is not. Warm and hot dust both seem to be independent of the star's age and spectral type; populations are found around A- to K-type stars, with ages from 10s of Myr to several Gyr \citep{Ertel2020}. Theoretical attempts to explain warm and hot dust are discussed in \mbox{Section \ref{subsec: theoryHotAndWarmDust}}.

\subsubsection{Conclusions from dust observations}
\label{subsec: obsResolvedImagesOfDust}

This section summarises our observational constraints on dust around main-sequence stars. \figRef{demographicsOfDebrisDisks} shows the locations, luminosities and ages of cold and warm debris disks, from a sample of 179 systems\footnote{The sample and data are from \citet{Pearce2022ISPY} and references therein, including unpublished ALMA data generously shared by Luca Matr\`{a}.}. The extrasolar disks have typical luminosities of $10^{-2}$ to ${10^{-5}}$ times those of their stars, meaning these disks are much brighter than our Asteroid Belt and Kuiper Belt (both around ${10^{-7}}$). This is because modern technology is unable to detect dust as faint as ours around other stars. The figure shows that many disks lie at tens or hundreds \mbox{of au} from their stars, and that some resolved disks are very broad, and others very narrow. Debris-disk systems have ages from about ${10\myr}$ to several Gyr, and disks around older stars are generally fainter; this is consistent with the idea that debris disks lose mass over time.

\begin{figure}[h]
\centering
\includegraphics[width=16.5cm]{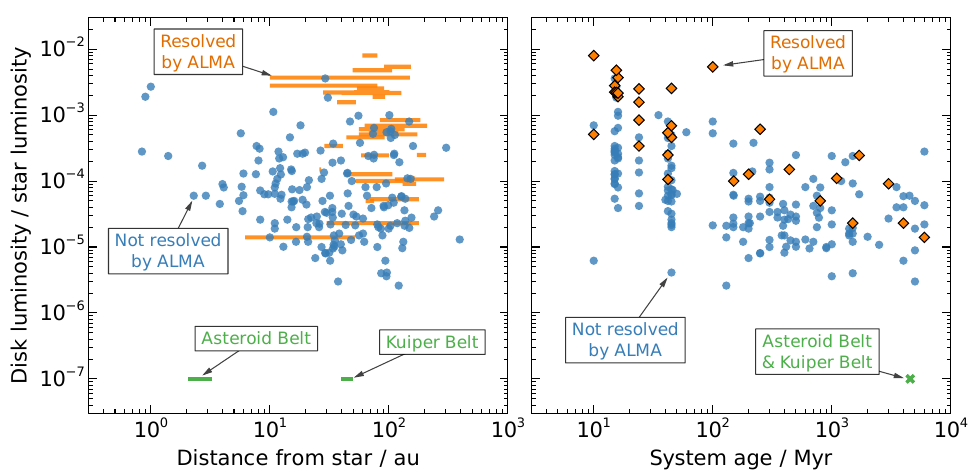}
\caption{Locations, luminosities and ages of cold and warm debris disks. On the left panel, orange bars show the widths of ALMA-resolved disks. Other disks are shown as blue points; the widths of these disks are often unconstrained from observations, so blue points show the approximate locations of the disk centres. The right panel shows the decline in disk brightness with age. The Asteroid Belt and Kuiper Belt are shown on both panels, and are much less luminous than detected extrasolar disks.}
\label{fig: demographicsOfDebrisDisks}
\end{figure}

Debris is found around main-sequence stars of all spectral types. Many of the brightest and most-studied debris disks are around A-type stars, because these massive stars are bright and often young, which makes debris easier to detect. Disks are also found around many intermediate, FGK-type stars like our Sun. Very few disks are found around M-type stars; this may be because these low-mass stars have fainter disks, which would make them harder to detect \citep{Luppe2020}. There may be a trend for disks around more luminous stars to have larger radii \citep{Matra2018}.

Resolved images show that many debris disks are not just symmetric, uniform rings. Instead, they have very diverse shapes and structures. Some disks are narrow and eccentric, like Fomalhaut on \figRef{fomalhautAtDifferentWavelengths}. Others are broad and contain gaps, like ${\HD107146}$ (\figRef{galleryOfDisks}\textit{a}). Some disks have warps, where the inner and outer regions are misaligned (like \mbox{\textbeta\;Pictoris}, shown edge-on on \figRef{galleryOfDisks}\textit{e}). Some have \glossaryTerm{clumps}, which are overdensities that do not extend all the way around the star. Some disks have sharp edges, whilst others have smoother edges \citep{ImazBlanco2023}. However, there are no clear detections of spirals in debris disks, which contrasts with the many spiral detections in \glossaryTerm{protoplanetary disks} (the young, massive disks of gas and dust from which debris and planets form). Finally, most resolved, cold disks are quite flat; disks seen edge-on in millimetre wavelengths typically have vertical thicknesses that are a few percent of their radii.

Most debris disks do not vary between observations, but there are some exceptions. Some disks with unusually high infrared excesses, dubbed \glossaryTerm{extreme debris disks}, undergo large brightness changes (e.g. \citealt{Meng2015}). These may be caused by collisions between large objects during terrestrial-planet formation. Some resolved disks are also variable, such as \mbox{AU Microscopii}, which hosts features seen to move away from the star over several years. Two such features are visible on \figRef{galleryOfDisks}\textit{d}, as the two bumps above the left side of the disk. 

Studies have also searched for correlations between debris disks and various system parameters. Some find that debris is more common in systems with low-mass planets, and less common in systems with high-mass planets, although other studies find no such correlations (e.g. \citealt{Yelverton2020}). Studies also analysed the \glossaryTerm{metallicity} of debris-disk stars, which means the abundance of elements other than hydrogen and helium. No clear correlations between star metallicity and debris are identified, although fewer disks may be detected around lower-metallicity stars (e.g. \citealt{Gaspar2016}).

\subsection{Gas observations}
\label{subsec: obsGas}

Dust is not the only detectable component of debris disks; we also see gas. Some gas is co-located with known dust belts, and gas at other locations appears to be released by comets. Both gas populations are discussed below.

\subsubsection{Gas near dust belts}

Gas in debris disks can be identified through \glossaryTerm{spectral lines}, which are sharp features in SEDs at very specific wavelengths. Examples of spectral lines can be seen on \figRef{betaPicSED}, as the small spikes in the star's SED. These correspond to specific quantum processes, such as emission when an electron moves to a lower energy level in an atom. Spectral lines can identify a specific atom or molecule, and the lines on \figRef{betaPicSED} are associated with processes in the star. However, gas in debris disks also has spectral lines, and these are used to detect the gas\footnote{Spectral lines are also used to probe dust compositions. For example, dust made of silicates has a spectral line around ${10\um}$, which can be identified in SEDs.}.

The gas most commonly considered in debris disks is carbon monoxide, or CO. This is because CO has strong spectral lines that are relatively easy to detect, located at wavelengths around ${1\mm}$. These are emission lines, meaning they correspond to spikes in the SED (rather than dips). \figRef{betaPicGas} shows an ALMA image of CO gas in the \mbox{\textbeta\;Pictoris} system (right panel), compared to a \glossaryTerm{continuum} image of millimetre-sized dust (left panel). \glossaryTerm{Continuum} means the component of an SED that is relatively smooth, as opposed to the sharp features at spectral lines.

\begin{figure}[h]
\centering
\includegraphics[width=16.5cm]{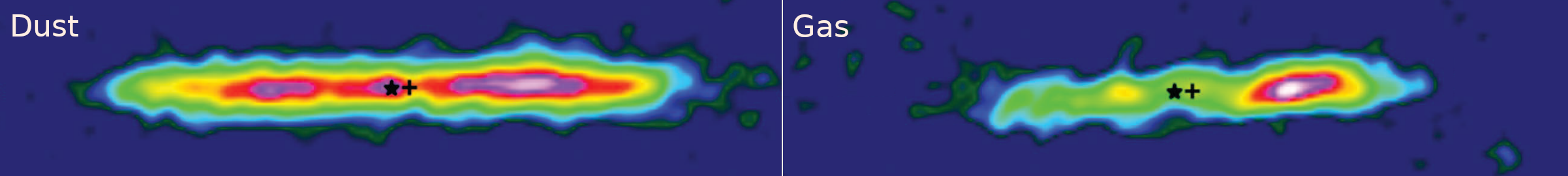}
\caption{Dust and gas in the \mbox{\textit{\textbeta} Pictoris} system, imaged by ALMA \citep{Dent2014}. The left panel shows millimetre-sized dust, and the right panel shows CO gas. The star and plus symbols mark the star and a detected planet, respectively. The disk is seen edge on, and each panel has the same scale and spans about \mbox{350 au} from left to right. The gas lies at a similar location to the dust, and there is a clump in the gas on the right side of the image.
}
\label{fig: betaPicGas}
\end{figure}

CO is detected in a few tens of nearby debris disks, which tend to be in young systems \citep{Hughes2018}. Most CO detections are around A-type stars, but there are some around Sun-like stars too. The amount of gas in debris disks is significantly less than in protoplanetary disks, and the origin of debris-disk gas is unclear. One possibility is that CO in debris disks is released during collisions between debris bodies with a high \glossaryTerm{volatile} content, where \glossaryTerm{volatiles} are substances that easily vaporize. Another possibility is that the gas is primordial, left over from the gas-rich protoplanetary disk.

Gas teaches us about the composition of debris. Using CO observations and collisional arguments, it was shown that planetesimals in several extrasolar debris disks contain similar quantities of CO and CO$_2$ as Solar System comets (e.g. \citealt{Matra2017}). Gas can also teach us about disk rotation. For a rotating disk, light from the part rotating away from us is slightly redshifted, and light from the part rotating toward us is slightly blueshifted. Since spectral lines occur at specific wavelengths, we can use them to measure the redshift at different points in a gas disk. This yields the disk's rotation direction (e.g. \citealt{Matra2017}).

\subsubsection{Extrasolar comets}

Gas also lets us detect exocomets around other stars. As a comet approaches a star, it heats up and releases gas. If this gas lies between us and the star, then we may detect the gas in absorption. This means we see dips in the star's SED, corresponding to spectral lines in the gas. Some absorption lines vary with time, often over hours, which implies that we detect \textit{individual} comets as they pass close to stars.

One of the most-studied exocomet systems is \mbox{\textbeta\;Pictoris}, which has very frequent and significant absorption events. The comets responsible for this absorption appear to have \glossaryTerm{periastron} distances of order ${0.1\au}$, where \glossaryTerm{periastron} means the point on an orbit that is closest to the star. Surveys detect variable absorption indicative of comets around about 14\% of stars, and there may be a tentative correlation between this variable absorption and hot dust \citep{Rebollido2020}. 

\horizontalLine

\section{Debris-disk theory}
\label{sec: theory}

The first part of the chapter summarised debris-disk observations. This second part summarises our theoretical understanding of debris disks. \mbox{Section \ref{subsec: theoryOriginOfDebrisDisks}} outlines how we think debris disks form, and \mbox{Section \ref{subsec: theoryWhyDifferentAtDifferentWavelengths}} explains why they look different at different wavelengths. \mbox{Section \ref{subsec: theorySizeDistAndCollCascade}} describes how dust is produced in debris disks, and what this implies about larger debris bodies. \mbox{Section \ref{subsec: theoryPlanetDebrisInteractions}} details how planets interact with debris, and \mbox{Section \ref{subsec: theoryHotAndWarmDust}} describes our efforts to understand warm and hot dust.

\subsection{Origins of debris disks}
\label{subsec: theoryOriginOfDebrisDisks}

Stars form when a giant cloud of gas and dust collapses under its own gravity. As it collapses, the cloud fragments into smaller parts. Each part becomes hotter and denser, and eventually forms one or more protostars at its core. Some material has too much angular momentum to fall directly onto the core, and this material collapses down into a disk around the protostar(s). This is the protoplanetary disk. After a few Myr, the protoplanetary disk disperses. The actual dispersion process is unknown, but it seems to be relatively fast. Some observed disks may currently be in this dispersion phase, such as \glossaryTerm{transition disks} and \glossaryTerm{hybrid disks}, but this is unclear.

After dispersion, much of the gas and dust from the original protoplanetary disk has gone. It may have formed into planets, accreted onto a star, or been blown away. What is left is a young planetary system, containing one or more stars, planets, planetesimals, and perhaps small quantities of gas and dust. The planetesimals, which can be thought of as remnant bodies left over from planet formation, comprise the young debris disk. Over time these planetesimals collide with each other, releasing the dust that we see (\mbox{Section \ref{subsec: theorySizeDistAndCollCascade}}).

\subsection{Why do debris disks look different at different wavelengths?}
\label{subsec: theoryWhyDifferentAtDifferentWavelengths}

\figRef{fomalhautAtDifferentWavelengths} showed that debris disks look different at different wavelengths. This is because different wavelengths probe different grain sizes, and different grain sizes have different spatial distributions. This section examines the physical reasons behind this.

\begin{BoxTypeA}[chap1:box1]{Key concept: \boxTextColour{different bodies experience different forces}}

\noindent Larger debris, such as asteroids and dwarf planets, are predominantly affected by gravity. Smaller debris, such as dust, are also affected by non-gravitational forces. This causes different debris sizes to have different spatial distributions, and so debris disks look different at different wavelengths. The distribution of larger grains tends to be more confined than that of smaller grains.

\end{BoxTypeA}

\subsubsection{Forces on debris}
\label{subsec: theoryForcesOnAGrain}

Debris orbiting a star is subject to several forces, as well as the inward pull of the star's gravity. These additional forces are described below.

\paragraph{Radiation pressure}

\glossaryTerm{Radiation pressure} arises due to photons from the star hitting a dust grain, and this force acts in the opposite direction to gravity. Since both gravity and photon flux scale with $r^{-2}$, where $r$ is the distance to the star, the main effect of radiation pressure is to reduce the net inward pull on the grain. The gravitational force is

\begin{equation}
    \fGrav = -\frac{G \mStar \mDust}{r^2} \boldsymbol{\hat{r}},
    \label{eq: gravity}
\end{equation}

\noindent where $G$, $\mStar$ and $\mDust$ are the gravitational constant, star mass and grain mass respectively, and $\boldsymbol{\hat{r}}$ is the unit vector in the star-grain separation direction. For a grain on a circular orbit,  the combined effect of radiation pressure and gravity produces a net radial force of

\begin{equation}
    \boldsymbol{F}_{\boldsymbol r} = \fGrav(1-\beta),
    \label{eq: gravityWithRadiationPressure}
\end{equation}

\noindent where $\beta$ is a positive value denoting the ratio of radiation pressure to gravity \citep{Burns1979}. The $\beta$ value depends on grain size, shape, composition and star type, with smaller grains generally having larger $\beta$ values. This means that smaller grains are usually more affected by radiation pressure than larger grains.

To show the effect of radiation pressure, consider a body on a circular orbit, which releases two dust grains. One grain has ${\beta=0}$, and the other has ${\beta > 0}$. The grain with ${\beta=0}$ feels only gravity, so would enter a circular orbit. Conversely, the grain with ${\beta > 0}$ would enter a non-circular orbit, with its furthest point (\glossaryTerm{apoastron}) outside the release point. If $\beta$ were high enough, then radiation pressure would overcome gravity, and the grain would be unbound and blow out of the system. This occurs if the grain is smaller than the \glossaryTerm{blowout size}. For grains released from circular orbits, blowout occurs if ${\beta \geq 0.5}$, which corresponds to a blowout size of ${1\um}$ for Sun-like stars and ${10\um}$ for more-luminous A0-type stars.

\begin{BoxTypeA}[chap1:box1]{Key concept: \boxTextColour{the smallest dust is around the blowout size}}

\noindent Radiation pressure is stronger for smaller grains, and grains smaller than the blowout size are usually blown away. Hence the smallest dust in a debris disk is usually around the blowout size.

\end{BoxTypeA}

\paragraph{Poynting-Robertson (P-R) drag}

Like radiation pressure, \glossaryTerm{Poynting-Robertson drag} (\mbox{P-R drag}) is caused by photons from the star hitting the dust grain. An orbiting grain has some non-radial velocity component, so the grain `sees' photons arriving at a slight angle, with a small velocity component opposing the grain's motion. This results in a drag force. The combined effects of gravity, radiation pressure and \mbox{P-R drag} result in a net force

\begin{equation}
    {\color{black}\boldsymbol{F} \; = \; \;}_
    {}
    {\color{red} \underbrace{\color{black} \fGrav(1-\beta)}_
    {\mathclap{\color{red}\text{Gravity + rad. pressure}}}}
    {\; \; - \; \;}
    {\color{red} \underbrace{\color{black} \beta |\fGrav| \left(\frac{\dot{r}_{\rm d}}{c} \rHat \; + \; \frac{\boldsymbol{v}_{\rm d}}{c}\right)}_
      {\color{red} \text{P-R drag}}}
    \label{eq: gravityAndRadiationForces}
\end{equation}

\noindent where $\boldsymbol{v}_{\rm d}$ is the dust velocity, $\dot{r}_{\rm d}$ is the radial component of $\boldsymbol{v}_{\rm d}$, and $c$ is the speed of light \citep{Burns1979}. The first term is the combination of gravity and radiation pressure (\eqRef{gravityWithRadiationPressure}), and the second term is \mbox{P-R drag}.

\mbox{P-R drag} has two main effects. It causes grains to spiral inwards towards the star, and causes eccentric orbits to become more circular. The timescales for \mbox{P-R drag} are generally much longer than orbital periods; for example, a grain with ${\beta=0.1}$, released at ${1\au}$ from the Sun, would take \mbox{4000 years} to spiral down to the Solar radius.

\paragraph{Stellar winds}

The effects of stellar winds are similar to radiation pressure and \mbox{P-R drag}. Winds reduce the net inward pull on grains, make them spiral inwards, and circularise orbits. The combined force of stellar winds plus gravity has a very similar form to \mbox{Equation \ref{eq: gravityAndRadiationForces}}, only with $\beta$ replaced by the ratio of the wind-pressure force to gravity, and $c$ replaced by the wind speed.

\paragraph{Other forces}

Several other forces also affect debris. These include the Lorentz force, which affects charged grains as they move through magnetic fields. Another force is gas drag, which can be important for small grains embedded in gaseous disks. There are also other radiation forces, which affect spinning or asymmetric bodies. Gravitational interactions with planets can also affect debris orbits, as detailed in \mbox{Section \ref{subsec: theoryPlanetDebrisInteractions}}.

\subsubsection{Forces on different grain sizes}

The forces in the previous section are the reason why debris disks have different shapes at different wavelengths. The strengths of these forces vary with grain size, so grains of different sizes have different orbits. This means that different grain sizes have different spatial distributions, which is why debris disks look different at different wavelengths.

Large grains, with sizes of millimetres to centimetres, essentially only feel the force of gravity. They are much less affected by other forces than small grains are. The orbits of large grains should be similar to those of even larger debris bodies, which would also move predominantly under gravity. This is why resolved observations of cold dust at millimetre wavelengths, which detect large grains, are thought to trace the distribution of larger, unseen planetesimals.

Unlike large grains, small grains are strongly affected by other forces. Grains smaller than a millimetre are susceptible to radiation or wind pressure, so their orbits can be highly eccentric or even unbound. This means that debris disks are often more extended at shorter wavelengths, which probe smaller grains. The halos in some systems are small grains on eccentric orbits, which were released from collisions in a smaller planetesimal belt (e.g. \figRef{galleryOfDisks}\textit{c}). These grains acquired wide orbits due to their large $\beta$ values. As well as halos, small grains can also extend inwards of their release locations, because P-R and wind drag make them spiral inwards. This inward migration is one possible way to supply exozodis (see \mbox{Section \ref{subsec: theoryHotAndWarmDust}}).

\subsection{Size distribution and the collisional cascade}
\label{subsec: theorySizeDistAndCollCascade}

Modern instruments can only detect dust and gas in extrasolar debris disks. Bodies larger than dust, like planetesimals and dwarf planets, are currently undetectable. However, we think that all debris disks contain such larger bodies, for two reasons. First, the Solar System's Asteroid and Kuiper Belts, which are thought to be similar to extrasolar disks, comprise a range of bodies from dust to dwarf planets. Second, dust cannot live as long as a star, yet a high fraction of stars have dust. This implies that grains get replenished somehow, and we think that replenishment occurs through collisions between planetesimals. This section summarises our understanding of collisions and the debris-size distribution.

\begin{BoxTypeA}[chap1:box1]{Key concept: \boxTextColour{dust in debris disks comes from violent collisions}}

\noindent The lifetimes of individual dust grains are much shorter than the ages of stars. For typical parameters, grains survive for around ${1\myr}$ before being destroyed by collisions, whilst the ages of debris-disk stars are tens or thousands of Myr. This means that the dust in debris disks cannot be left over from system formation. Instead, we think dust is continually replenished by collisions between larger bodies.

\end{BoxTypeA}

\subsubsection{Collisions}
\label{subsec: theoryCollisions}

When two bodies collide, there are several possible outcomes. The bodies could either merge, bounce, or break apart. The extreme is a \glossaryTerm{catastrophic collision}, where bodies undergo significant fragmentation, and the largest fragment has less than half the mass of the original body. Which outcome occurs depends on the collision velocity, and the strength and composition of the bodies.  

Dust in debris disks comes from the collisional fragmentation of planetesimals. This means that collisions must be violent enough to break planetesimals apart. The required collision speed can be predicted by considering the impact energy required for a catastrophic collision, $\QDStar$. This depends on the sizes of the bodies, and follows a V-shaped dependence as shown on \figRef{vFrag}. Large bodies like dwarf planets are held together by gravity; in this regime, the bigger the body, the stronger it is. Conversely, smaller bodies like dust are held together by material strength; in this regime, the \textit{smaller} the body, the stronger it is. Bodies are therefore strongest if they are very small or very large, and weakest at some intermediate size. \figRef{vFrag} shows that collision speeds in debris disks must be at least 10s or 100s of metres per second to release dust, although this depends on the debris composition. 

\begin{figure}[h]
\centering
\includegraphics[width=14cm]{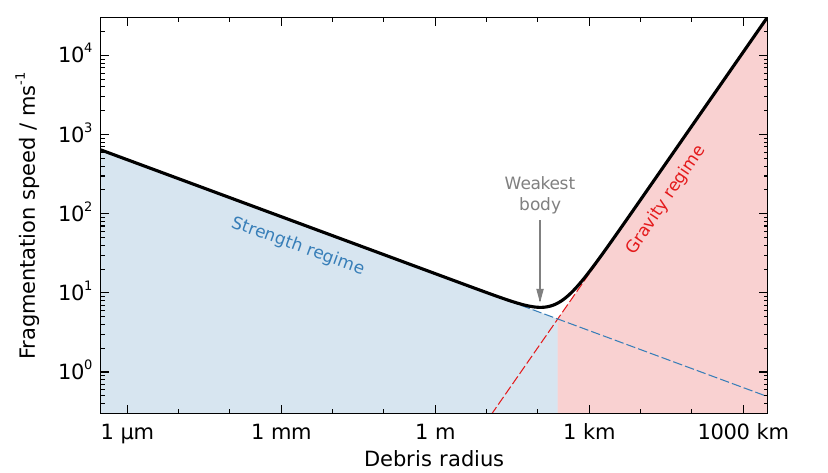}
\caption{Collision speed required to break up two equal-sized bodies made of basalt, adapted from \citet{Costa2024}. If the collision speed is above the black solid line, then a catastrophic collision occurs, resulting in significant fragmentation. Such a collision would produce a significant amount of dust.}
\label{fig: vFrag}
\end{figure}

\begin{BoxTypeA}[chap1:box1]{Key concept: \boxTextColour{debris disks lose mass and get fainter over time}}

\noindent Collisions grind debris into smaller and smaller sizes, which eventually blow out of the system or sublimate near the star. This means that debris disks lose mass over time. Also, once the largest bodies start colliding, the amount of dust starts dropping. This means that debris disks become fainter over time, so they are easier to detect around younger stars. This can be seen on the right panel of \figRef{demographicsOfDebrisDisks}, where older debris disks are generally fainter.

\end{BoxTypeA}

\subsubsection{Debris-size distribution}
\label{subsec: theorySizeDist}

Larger debris collides and breaks into smaller bodies, and those smaller bodies collide and break into even smaller bodies. This grinding of debris into ever-smaller bodies is called a \glossaryTerm{collisional cascade}. Given sufficient time, a collisional cascade reaches a steady state, and debris assumes a particular size distribution. This is approximated by the powerlaw

\begin{equation}
n(s) {\rm d}s \propto s^{-3.5} {\rm d}s,
\label{eq: dohnanyiSizeDist}
\end{equation}

\noindent where ${n(s){\rm d}s}$ is the number of bodies with radii between $s$ and ${s+{\rm d}s}$ \citep{Dohnanyi1969}. The steep index of ${-3.5}$ means that smaller bodies significantly outnumber larger bodies; for example, there would be 300 times as many bodies with radii between ${0.1}$ and ${1\mm}$ as there would be between ${1}$ and ${10\mm}$. The upper end of the size distribution is set by the largest bodies, and the lower end is set by the blowout size. This size distribution is shown on \figRef{sizeDist}.

\begin{BoxTypeA}[chap1:box1]{Key concept: \boxTextColour{small debris dominates emission, large debris dominates mass}}

\noindent Bodies in a debris disk have a steep size distribution, meaning that smaller bodies are much more numerous than larger bodies. The index of $-3.5$ in \eqRef{dohnanyiSizeDist} means that a disk's total surface area (and hence brightness) is dominated by the smallest grains, whilst its total mass is dominated by the largest bodies. 

\end{BoxTypeA}

Real size distributions would deviate from the simple powerlaw of \eqRef{dohnanyiSizeDist}. Realistic distributions would be complicated at small sizes by the effects of blowout, \mbox{P-R drag} and winds. Also, at large sizes, bodies would take longer to enter the collisional cascade and start breaking up, so the largest bodies may not follow a collisional size distribution until later in the system's life. However, despite these complications, we think that \eqRef{dohnanyiSizeDist} is a reasonable estimate of the debris-size distribution. 

\begin{figure}[h]
\centering
\includegraphics[width=14cm]{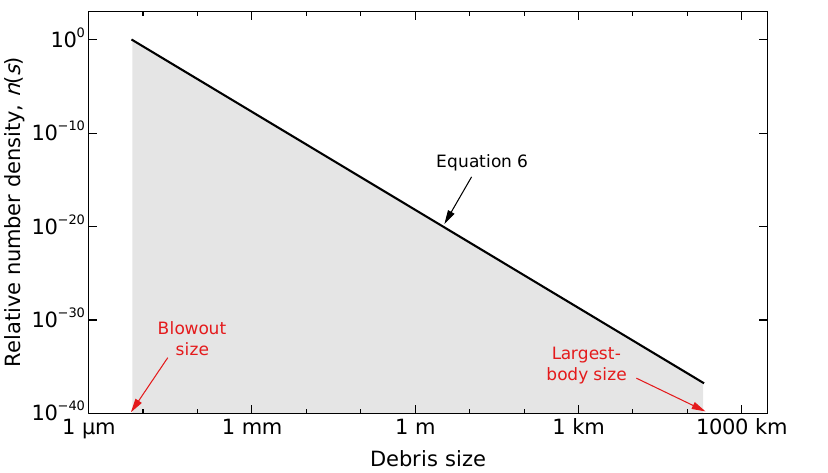}
\caption{Theoretical size distribution of bodies in a debris disk (\eqRef{dohnanyiSizeDist}).
Smaller bodies are much more numerous than larger bodies, and the smallest dust is at the blowout size. In this example the largest body is arbitrarily set to \mbox{200 km}, but we do not actually know the sizes of the largest bodies in debris disks. More-realistic models of size distributions have additional features, but most follow this rough shape.}
\label{fig: sizeDist}
\end{figure}

\subsubsection{Stirring}
\label{subsec: theoryStirring}

For collisions to release dust, debris orbits must intersect (for collisions to occur), and have high relative velocities (for collisions to be violent). This means that debris orbits must be excited, i.e. have significant eccentricities and/or mutual inclinations. However, we think that debris forms with unexcited orbits, due to conditions in the protoplanetary disk. This implies that some process then excites debris. The debris-excitation process is called \glossaryTerm{stirring}, and is thought to have happened in all observed debris disks.

We do not know how debris disks get stirred. Several stirring models have been proposed, but it is difficult to distinguish them using current observations, and each model has potential problems when considered in isolation. Current models include:

\vspace{1mm}

\begin{itemize}
    \item \textbf{Self-stirring}, where debris bodies excite each other. Self-stirring could operate in some disks, but others would need implausibly high masses for this mechanism to work \citep{Krivov2021}.

    \vspace{1mm}
    
    \item \textbf{Planetary stirring}, where debris gets excited by dynamical interactions with planets. However, planet stirring could be inhibited by the self gravity of massive debris disks.

    \vspace{1mm}
        
    \item \textbf{Flyby stirring}, where another star passes close to the system and excites debris. However, flybys do not seem common enough to explain the high fraction of stars with observed dust \citep{Kenyon2002}.

    \vspace{1mm}

    \item \textbf{Pre-stirring}, where planetesimals are already stirred in the protoplanetary disk. One way this could happen is if planetesimals formed on excited orbits, but this process is currently unclear. 
    
\end{itemize}

\vspace{1mm}

\subsection{Planet-debris interactions}
\label{subsec: theoryPlanetDebrisInteractions}

So far, the theory section has mainly described processes occurring within debris disks. However, debris does not exist in isolation. Many planetary systems host both planets and debris, and these objects can gravitationally interact with each other. For example, Neptune dominates the dynamical evolution of Kuiper Belt objects, and Jupiter carves gaps in the Asteroid Belt. Similar interactions should occur in extrasolar systems, and we even see it in some cases. For example, we detect a planet orbiting \mbox{\textbeta\;Pictoris} which is probably responsible for that disk's warp.

Many extrasolar debris disks have features that imply planetary interactions, such as gaps, sharp edges, and eccentric shapes. We usually cannot detect the planets responsible, because they would lie in the outer regions of systems, which is where planet-detection techniques struggle. However, we can use features in observed debris disks to infer the properties of unseen planets. \figRef{debrisInferredPlanets} shows the predicted population of unseen planets inferred from debris disks.

\begin{figure}[h]
\centering
\includegraphics[width=10cm]{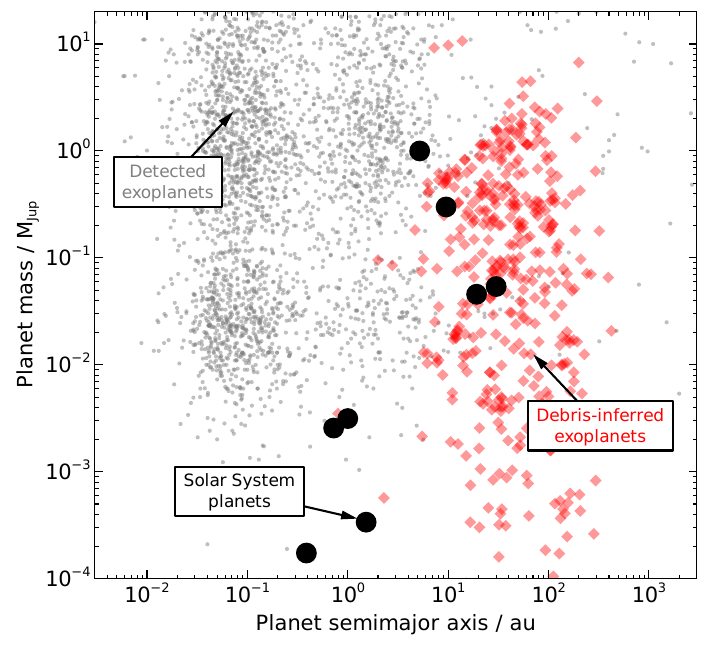}
\caption{Unseen exoplanets inferred from debris disks (red diamonds, from \citealt{Pearce2022ISPY}), compared to Solar System planets (black circles) and known exoplanets (grey dots, from \mbox{\href{https://exoplanet.eu/home/}{exoplanet.eu}}, accessed \mbox{Feb 2024}). Some of the inferred planets may soon be detected by \textit{JWST}, which is sensitive to planets above \mbox{0.1 Jupiter} masses in the outer regions of systems.}
\label{fig: debrisInferredPlanets}
\end{figure}

This section outlines what happens in planet-debris interactions. There are three types of interaction: \glossaryTerm{scattering} interactions, \glossaryTerm{secular} interactions, and \glossaryTerm{mean-motion resonances}. \figRef{planetDiskInteractionsSchematic} shows a typical interaction between an eccentric planet and a broad, external debris disk. Three debris populations are highlighted, each undergoing a different type of interaction. The effect of each interaction on debris is explained below.

\begin{figure}[h]
\centering
\includegraphics[width=16.5cm]{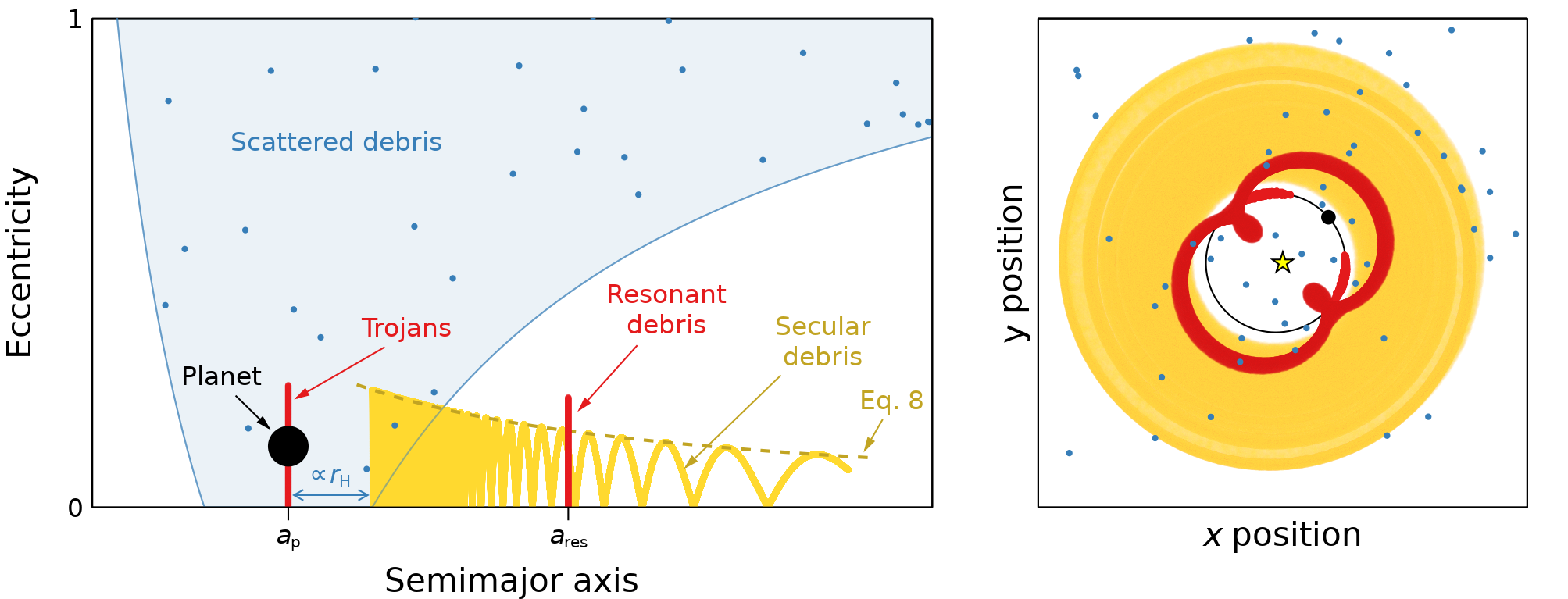}
\caption{Diagram of a typical planet-debris interaction. In this example the planet's orbit is eccentric, and the disk's mass is negligible. The left panel shows orbital eccentricities and semimajor axes, and the right panel shows positions. Debris is coloured by interaction on both panels: scattered debris is blue, secular is yellow, and resonant is red. On the right panel, the star is at the centre, the planet's orbit is the black line, and the planet is the black circle. Only two resonant populations are shown, but in reality there can be debris in many different MMRs, each with different shapes on the right panel.}
\label{fig: planetDiskInteractionsSchematic}
\end{figure}

\paragraph{Scattering interactions}

Scattering happens when two bodies pass near each other. This is a rapid, short-range interaction, which changes the semimajor axes, eccentricities and orientations of orbits. Scattering interactions are often invoked to explain gaps and sharp edges in debris disks, and scattering of debris by planets can also drive planet migration (e.g. \citealt{Friebe2022}).

Small bodies get scattered if they pass within a few \glossaryTerm{Hill radii} of a planet, which is the region where the planet's gravity dominates over that of the star. For a planet on a circular orbit, the Hill radius is

\begin{equation}
    \rHill = \aPlt \left(\frac{\mPlt}{3 \mStar}\right)^{1/3},
    \label{eq: hillRadius}
\end{equation}

\noindent where $\mPlt$ and $\aPlt$ are the planet mass and semimajor axis, respectively. Scattered debris is shown in blue on \figRef{planetDiskInteractionsSchematic}. On the left panel, the blue region shows orbits passing within several Hill radii of the planet. Scattered bodies are confined to this region, jumping from one orbit to another with each scattering event. On the right panel, scattered debris forms a diffuse cloud. Bodies may get repeatedly scattered, and eventually ejected from the system.

\paragraph{Secular interactions}

Secular interactions occur over timescales much longer than orbital periods. They change the eccentricities and orientations of orbits, so are often invoked to explain eccentric or warped debris disks. An eccentric planet would make external debris oscillate in eccentricity. For a low-eccentricity planet interacting with coplanar debris, debris eccentricity would oscillate up to a maximum

\begin{equation}
    \eMax \approx \frac{5}{2} \frac{\aPlt}{a} \ePlt,
    \label{eq: twoEForced}
\end{equation}

\noindent where $a$ is the debris semimajor axis and $\ePlt$ is the planet eccentricity. The debris orientation would also evolve, which results in an eccentric disk aligned with the planet's orbit. This is visible on \figRef{planetDiskInteractionsSchematic}, where secular debris is yellow. The waves on the left panel, and the spirals on the right panel, would tighten over time.

Secular interactions also occur if the planet is inclined, which makes debris inclinations oscillate. In case of extreme inclinations, the \glossaryTerm{von Zeipel–Kozai–Lidov mechanism} can occur, where orbits oscillate between having low eccentricity with high inclination, and high eccentricity with low inclination.

\paragraph{Mean-motion resonances (MMRs)}

Mean-motion resonances (MMRs) occur if orbital periods are close to simple multiples of each other. For example, Neptune and Pluto are in the \mbox{3:2 MMR}, meaning that Neptune completes three orbits for every two of Pluto. MMRs can keep orbits stable; Pluto's orbit overlaps Neptune's, but the \mbox{3:2 MMR} ensures the bodies never come close enough to scatter. 
MMRs can also make bodies unstable, by making gravitational perturbations accumulate; for example, some gaps in Saturn's rings arise through MMRs with its moons.

MMRs are denoted using the notation \mbox{$(p+q):p$}, so for example the \mbox{3:2 MMR} has ${p=2}$ and ${q=1}$. The \glossaryTerm{nominal}, or central, location of an MMR is derived from the ratio of orbital periods. For an MMR outside a planet's orbit, this is

\begin{equation}
    \aRes = \aPlt \left(\frac{p+q}{p}\right)^{2/3},
    \label{eq: mmrLocation}
\end{equation}

\noindent where $\aRes$ is the nominal semimajor axis of the resonance. Generally, the closer a body's semimajor axis is to $\aRes$, the greater the effect of the resonance. The effect also depends on the \glossaryTerm{order} of the resonance, $q$, where lower $q$ usually means a stronger effect.

Resonant debris is shown in red on \figRef{planetDiskInteractionsSchematic}. Two populations are shown: \glossaryTerm{Trojans} in the 1:1 MMR, which share the planet's orbit, and another population with nominal semimajor axis $\aRes$. Resonant debris oscillates in eccentricity, whilst its semimajor axis stays close to the nominal location (left panel). This debris forms distinct structures, which are symmetric around the planet's position (right panel). MMRs are often used to explain clumps in debris disks, such as the possible clumps in \mbox{\textepsilon\;Eridani} (\figRef{galleryOfDisks}\textit{b}).

\subsection{Models of warm and hot dust}
\label{subsec: theoryHotAndWarmDust}

This last section specifically discusses warm and hot dust. Some warm dust may be released from planetesimal belts near the habitable zone, like the Asteroid Belt in the Solar System. However, some systems have too much warm dust to be compatible with this model. This is because warm dust is much closer to stars than cold dust, so its collisional evolution is much faster. This would cause planetesimal disks with sufficient mass at these locations to quickly grind away to nothing, so local collisional cascades cannot be the source of all warm dust. Hot dust is even closer to stars than warm dust, so this problem is even more acute. We therefore think that other mechanisms are responsible for hot dust, as well as some populations of warm dust.

Warm dust may originate further out in the system, then get transported inwards to the habitable zone. Several mechanisms have been proposed to do this. In one model, dust from a distant planetesimal disk migrates inwards through \mbox{P-R} or wind drag \citep{Reidemeister2011}. In another model, warm dust is released from comets as they travel through the habitable zone, which is probably how Zodiacal dust is replenished in the Solar System \citep{Bonsor2014}. In a third model, warm dust is released by a single giant collision, rather than a steady state collisional cascade. Each of these models can explain observed warm dust, without requiring massive planetesimal disks in the habitable zone.

Hot dust is harder to explain. This appears to be very small, very hot grains, located very close to stars \citep{Kirchschlager2017}. However, these grains should rapidly sublimate or blow away, so we do not know how these populations survive. Many models have been proposed, but none are completely successful. Again, we think that hot dust originates further out and gets transported inwards, but the above models for warm dust do not work for hot dust. Dust migration via \mbox{P-R drag} is incompatible with hot-dust observations (e.g. \citealt{vanLieshout2014}), and cometary supply would require unphysically high inflow rates \citep{Pearce2022Comets}. Instead, some mechanism may trap hot dust near the star. This could be stellar magnetic fields (e.g. \citealt{Kimura2020}), or gas released by sublimating grains \citep{Pearce2020}, but both models have problems. Hot dust is an area of significant research, but it currently remains a mystery.

Warm and hot dust could hinder searches for Earth-like exoplanets. Soon, new facilities will try to image Earth-like planets in the habitable zone, and characterise their atmospheres. The hope is to assess the planets' potential for life. However, dust could act as noise, producing emission and making planet observations difficult. Therefore, there is considerable interest in understanding warm and hot dust, partly to mitigate its effect on future exo-Earth searches.

\horizontalLine

\section{Unsolved problems in debris-disk science}
\label{sec: unsolvedProblems}

We have made substantial progress in understanding debris disks over the past few decades. However, there are still many unanswered questions. This final part of the chapter lists some of the biggest questions, which are all active areas of research.

\paragraph{How massive are debris disks?}

Currently, we cannot detect anything larger than a centimetre in extrasolar debris disks. This means we do not know how big, or numerous, the largest bodies are. We therefore do not know the masses of extrasolar disks, because disk masses should be dominated by the largest bodies (\mbox{Section \ref{subsec: theorySizeDist}}). Several studies attempted to estimate disk mass by extrapolating the mass in observed dust, which requires assumptions about the size distribution and the largest-body size. However, these inferred masses are often unrealistically large, which is an issue known as the debris-disk-mass problem \citep{Krivov2021}. The solution to this problem could be that the largest bodies are much smaller than those in the Solar System, or that the actual size distribution is different to our models. Either way, determining the masses of debris disks would help us understand their origin, evolution, and how they interact with planets.

\paragraph{How are debris disks stirred?}

We often assume that debris must be stirred to induce destructive collisions. However, we do not know how this happens, or whether different disks get stirred by different processes. The idea that debris needs stirring at all is also coming under scrutiny. Answering these questions would teach us about debris evolution, and the transition from protoplanetary disks to debris disks.

\paragraph{Where does cold gas come from?}

There are several plausible origins for the gas that is co-located with dust disks. Gas could be primordial, meaning it is left over from the protoplanetary disk, or secondary, meaning it is continually released. Since only some gas species are observed, like CO, we do not know how much gas exists in other species, like hydrogen. Primordial and secondary models predict very different ratios of observable to non-observable gas, so answering where gas comes from would help constrain how much is present. This has implications for debris composition and disk dynamics.

\paragraph{What forms debris-disk structures?}

Debris disks have diverse shapes and features (\figRef{galleryOfDisks}). This diversity is sometimes ascribed to planets, but there are also alternative explanations. For example, debris structures could result from stellar flybys, or reflect where solids form in protoplanetary disks. Answering why debris disks are diverse would teach us about the architectures, formation and evolution of planetary systems.

\paragraph{Are debris disks connected to planets?}

It has long been thought that debris disks interact with unseen planets (\mbox{Section \ref{subsec: theoryPlanetDebrisInteractions}}). Now, \textit{JWST} offers the first real chance to detect these planets, and finding them would support our current ideas about disk sculpting and stirring. Conversely, not finding planets could imply that debris and planets are less linked than currently thought, and could prompt a rethink about why disks have the shapes and features that they do.

\paragraph{What is hot dust?}

No model has fully explained how hot dust survives in the face of sublimation and blowout, especially across a wide range of star ages and spectral types. Determining its nature would help us understand the innermost regions of planetary systems, and how material flows across systems. It would also help us mitigate the effect of dust on future exo-Earth imaging.

\horizontalLine

\section{Conclusions}
\label{sec: conclusions}

Debris is one of the major constituents of planetary systems. As instruments and techniques improve, we are detecting and imaging debris around more and more stars. We are also learning how debris evolves, where it comes from, and how it interacts with planets. The field of debris disks is a relatively small but growing area of astrophysics, and has potential to teach us about the formation, evolution and architecture of planetary systems.

\seealso{Several excellent reviews cover debris-disk observations and theory in much more detail. These include \cite{Hughes2018}, \cite{Krivov2010}, \cite{Marino2022}, \cite{Matthews2014}, \cite{Wyatt2008} and \mbox{\cite{Wyatt2020}}.}

\begin{ack}[Acknowledgments]
\phantom{} I thank Mark Booth, Meredith Hughes, Grant Kennedy, Alexander Krivov and Sebastian Marino for their expert advice and suggestions. I also appreciate the feedback on readability and understandability from Allan Cheruiyot, Abigail Falk, Jonathan Jackson, Jamar Kittling, Junu Lee, Elias Mansell, Yamani Mpofu, Aliya Nurmohamed, Francesca Pearce, Cat Sarosi and Brianna Zawadzki. I am grateful for support from a Warwick Prize Fellowship, made possible by a generous philanthropic donation, and from UKRI/EPSRC through a Stephen Hawking Fellowship.
\end{ack}

\horizontalLine

\bibliographystyle{Harvard}
\bibliography{bib}

\end{document}